\documentstyle[12pt,epsfig]{article}
\def\kev{\mbox{keV }}

\def\gevc{\mbox{ GeV/$c$}}
\def\gevsq{\mbox{ GeV/$c^{2}$}}
\def\mevsq{\mbox{ MeV/$c^{2}$}}
\def\dedx{\mbox{d{\it E}/d{\it x }}}

\def\pb{${\mathrm pb^{-1}}$ }

\def\CL{\mbox{C.L. }}

\def\Jpsi{\mbox{J/$\psi$ }}

\begin{document}
\begin{center}
\large EUROPEAN LABORATORY FOR PARTICLE PHYSICS (CERN)
\end{center}
\vspace{1cm}
\begin{flushright}
CERN-EP/99-153\\
22 October 1999
\end{flushright} 
\vspace{2.5cm}
\begin{center}
\Large {\bf Search for the glueball candidates 
{\boldmath $f_{0}(1500)$} and {\boldmath $f_{J}(1710)$}
 in {\boldmath $\gamma \gamma$} collisions}\\ 
\vspace{2cm}
\normalsize 
\large The ALEPH Collaboration\footnote[1]{See next page for the list of 
authors}
\vspace{1.5cm}
\end{center}
\normalsize
\begin{abstract}
  Data taken with the ALEPH detector at LEP1 have been 
 used to search for $\gamma\gamma$ production of the glueball 
 candidates $f_{0}(1500)$ and $f_{J}(1710)$ via their decay 
to $\pi^{+}\pi^{-}$. 
 No signal is observed and  
 upper limits to the product of $\gamma\gamma$ width and
$\pi^{+}\pi^{-}$ branching ratio of the $f_{0}(1500)$ 
 and the $f_{J}(1710)$ have been measured to be
$$\Gamma(\gamma \gamma \rightarrow f_{0}(1500)) \cdot {\cal BR}(f_{0}(1500)
 \rightarrow\pi^{+}\pi^{-}) < 0.31 \ \kev $$ 
\begin{center} and \end{center} 
$$\Gamma(\gamma \gamma \rightarrow f_{J}(1710)) \cdot {\cal BR}(f_{J}(1710)
 \rightarrow \pi^{+}\pi^{-}) <  0.55 \ \kev $$
at 95\% confidence level.
\end{abstract}
\vspace{2cm}
\begin{center}
\large \em{Submitted to Phys. Lett. B}
\end{center}
\normalsize
%
%
\pagestyle{empty}
\newpage
\small
%
\newlength{\saveparskip}
\newlength{\savetextheight}
\newlength{\savetopmargin}
\newlength{\savetextwidth}
\newlength{\saveoddsidemargin}
\newlength{\savetopsep}
\setlength{\saveparskip}{\parskip}
\setlength{\savetextheight}{\textheight}
\setlength{\savetopmargin}{\topmargin}
\setlength{\savetextwidth}{\textwidth}
\setlength{\saveoddsidemargin}{\oddsidemargin}
\setlength{\savetopsep}{\topsep}
%
%
\setlength{\parskip}{0.0cm}
\setlength{\textheight}{25.0cm}
\setlength{\topmargin}{-1.5cm}
\setlength{\textwidth}{16 cm}
\setlength{\oddsidemargin}{-0.0cm}
\setlength{\topsep}{1mm}
\pretolerance=10000
\centerline{\large\bf The ALEPH Collaboration}
\footnotesize
\vspace{0.5cm}
{\raggedbottom
\begin{sloppypar}
\samepage\noindent
R.~Barate,
D.~Decamp,
P.~Ghez,
C.~Goy,
\mbox{J.-P.~Lees},
E.~Merle,
\mbox{M.-N.~Minard},
B.~Pietrzyk
\nopagebreak
\begin{center}
\parbox{15.5cm}{\sl\samepage
Laboratoire de Physique des Particules (LAPP), IN$^{2}$P$^{3}$-CNRS,
F-74019 Annecy-le-Vieux Cedex, France}
\end{center}\end{sloppypar}
\vspace{2mm}
\begin{sloppypar}
\noindent
R.~Alemany,
S.~Bravo,
M.P.~Casado,
M.~Chmeissani,
J.M.~Crespo,
E.~Fernandez,
\mbox{M.~Fernandez-Bosman},
Ll.~Garrido,$^{15}$
E.~Graug\`{e}s,
A.~Juste,
M.~Martinez,
G.~Merino,
R.~Miquel,
Ll.M.~Mir,
A.~Pacheco,
I.~Riu,
H.~Ruiz
\nopagebreak
\begin{center}
\parbox{15.5cm}{\sl\samepage
Institut de F\'{i}sica d'Altes Energies, Universitat Aut\`{o}noma
de Barcelona, E-08193 Bellaterra (Barcelona), Spain$^{7}$}
\end{center}\end{sloppypar}
\vspace{2mm}
\begin{sloppypar}
\noindent
A.~Colaleo,
D.~Creanza,
M.~de~Palma,
G.~Iaselli,
G.~Maggi,
M.~Maggi,
S.~Nuzzo,
A.~Ranieri,
G.~Raso,
F.~Ruggieri,
G.~Selvaggi,
L.~Silvestris,
P.~Tempesta,
A.~Tricomi,$^{3}$
G.~Zito
\nopagebreak
\begin{center}
\parbox{15.5cm}{\sl\samepage
Dipartimento di Fisica, INFN Sezione di Bari, I-70126
Bari, Italy}
\end{center}\end{sloppypar}
\vspace{2mm}
\begin{sloppypar}
\noindent
X.~Huang,
J.~Lin,
Q. Ouyang,
T.~Wang,
Y.~Xie,
R.~Xu,
S.~Xue,
J.~Zhang,
L.~Zhang,
W.~Zhao
\nopagebreak
\begin{center}
\parbox{15.5cm}{\sl\samepage
Institute of High-Energy Physics, Academia Sinica, Beijing, The People's
Republic of China$^{8}$}
\end{center}\end{sloppypar}
\vspace{2mm}
\begin{sloppypar}
\noindent
D.~Abbaneo,
G.~Boix,$^{6}$
O.~Buchm\"uller,
M.~Cattaneo,
F.~Cerutti,
V.~Ciulli,
G.~Dissertori,
H.~Drevermann,
R.W.~Forty,
M.~Frank,
T.C.~Greening,
A.W. Halley,
J.B.~Hansen,
J.~Harvey,
P.~Janot,
B.~Jost,
I.~Lehraus,
O.~Leroy,
P.~Mato,
A.~Minten,
A.~Moutoussi,
F.~Ranjard,
L.~Rolandi,
D.~Schlatter,
M.~Schmitt,$^{20}$
O.~Schneider,$^{2}$
P.~Spagnolo,
W.~Tejessy,
F.~Teubert,
E.~Tournefier,
A.E.~Wright
\nopagebreak
\begin{center}
\parbox{15.5cm}{\sl\samepage
European Laboratory for Particle Physics (CERN), CH-1211 Geneva 23,
Switzerland}
\end{center}\end{sloppypar}
\vspace{2mm}
\begin{sloppypar}
\noindent
Z.~Ajaltouni,
F.~Badaud,
G.~Chazelle,
O.~Deschamps,
A.~Falvard,
C.~Ferdi,
P.~Gay,
C.~Guicheney,
P.~Henrard,
J.~Jousset,
B.~Michel,
S.~Monteil,
\mbox{J-C.~Montret},
D.~Pallin,
P.~Perret,
F.~Podlyski
\nopagebreak
\begin{center}
\parbox{15.5cm}{\sl\samepage
Laboratoire de Physique Corpusculaire, Universit\'e Blaise Pascal,
IN$^{2}$P$^{3}$-CNRS, Clermont-Ferrand, F-63177 Aubi\`{e}re, France}
\end{center}\end{sloppypar}
\vspace{2mm}
\begin{sloppypar}
\noindent
J.D.~Hansen,
J.R.~Hansen,
P.H.~Hansen,$^{1}$
B.S.~Nilsson,
B.~Rensch,
A.~W\"a\"an\"anen
\begin{center}
\parbox{15.5cm}{\sl\samepage
Niels Bohr Institute, DK-2100 Copenhagen, Denmark$^{9}$}
\end{center}\end{sloppypar}
\vspace{2mm}
\begin{sloppypar}
\noindent
G.~Daskalakis,
A.~Kyriakis,
C.~Markou,
E.~Simopoulou,
I.~Siotis,
A.~Vayaki
\nopagebreak
\begin{center}
\parbox{15.5cm}{\sl\samepage
Nuclear Research Center Demokritos (NRCD), GR-15310 Attiki, Greece}
\end{center}\end{sloppypar}
\vspace{2mm}
\begin{sloppypar}
\noindent
A.~Blondel,
G.~Bonneaud,
\mbox{J.-C.~Brient},
A.~Roug\'{e},
M.~Rumpf,
M.~Swynghedauw,
M.~Verderi,
H.~Videau
\nopagebreak
\begin{center}
\parbox{15.5cm}{\sl\samepage
Laboratoire de Physique Nucl\'eaire et des Hautes Energies, Ecole
Polytechnique, IN$^{2}$P$^{3}$-CNRS, \mbox{F-91128} Palaiseau Cedex, France}
\end{center}\end{sloppypar}
\vspace{2mm}
\begin{sloppypar}
\noindent
E.~Focardi,
G.~Parrini,
K.~Zachariadou
\nopagebreak
\begin{center}
\parbox{15.5cm}{\sl\samepage
Dipartimento di Fisica, Universit\`a di Firenze, INFN Sezione di Firenze,
I-50125 Firenze, Italy}
\end{center}\end{sloppypar}
\vspace{2mm}
\begin{sloppypar}
\noindent
M.~Corden,
C.~Georgiopoulos
\nopagebreak
\begin{center}
\parbox{15.5cm}{\sl\samepage
Supercomputer Computations Research Institute,
Florida State University,
Tallahassee, FL 32306-4052, USA $^{13,14}$}
\end{center}\end{sloppypar}
\vspace{2mm}
\begin{sloppypar}
\noindent
A.~Antonelli,
G.~Bencivenni,
G.~Bologna,$^{4}$
F.~Bossi,
P.~Campana,
G.~Capon,
V.~Chiarella,
P.~Laurelli,
G.~Mannocchi,$^{1,5}$
F.~Murtas,
G.P.~Murtas,
L.~Passalacqua,
\mbox{M.~Pepe-Altarelli}
\nopagebreak
\begin{center}
\parbox{15.5cm}{\sl\samepage
Laboratori Nazionali dell'INFN (LNF-INFN), I-00044 Frascati, Italy}
\end{center}\end{sloppypar}
\vspace{2mm}
\begin{sloppypar}
\noindent
L.~Curtis,
J.G.~Lynch,
P.~Negus,
V.~O'Shea,
C.~Raine,
\mbox{P.~Teixeira-Dias},
A.S.~Thompson
\nopagebreak
\begin{center}
\parbox{15.5cm}{\sl\samepage
Department of Physics and Astronomy, University of Glasgow, Glasgow G12
8QQ,United Kingdom$^{10}$}
\end{center}\end{sloppypar}
\vspace{2mm}
\begin{sloppypar}
\noindent
R.~Cavanaugh,
S.~Dhamotharan,
C.~Geweniger,$^{1}$
P.~Hanke,
G.~Hansper,
V.~Hepp,
E.E.~Kluge,
A.~Putzer,
J.~Sommer,
K.~Tittel,
S.~Werner,$^{19}$
M.~Wunsch$^{19}$
\nopagebreak
\begin{center}
\parbox{15.5cm}{\sl\samepage
Institut f\"ur Hochenergiephysik, Universit\"at Heidelberg, D-69120
Heidelberg, Germany$^{16}$}
\end{center}\end{sloppypar}
\vspace{2mm}
\begin{sloppypar}
\noindent
R.~Beuselinck,
D.M.~Binnie,
W.~Cameron,
P.J.~Dornan,
M.~Girone,
S.~Goodsir,
E.B.~Martin,
N.~Marinelli,
A.~Sciab\`a,
J.K.~Sedgbeer,
E.~Thomson,
M.D.~Williams
\nopagebreak
\begin{center}
\parbox{15.5cm}{\sl\samepage
Department of Physics, Imperial College, London SW7 2BZ,
United Kingdom$^{10}$}
\end{center}\end{sloppypar}
\vspace{2mm}
\begin{sloppypar}
\noindent
V.M.~Ghete,
P.~Girtler,
E.~Kneringer,
D.~Kuhn,
G.~Rudolph
\nopagebreak
\begin{center}
\parbox{15.5cm}{\sl\samepage
Institut f\"ur Experimentalphysik, Universit\"at Innsbruck, A-6020
Innsbruck, Austria$^{18}$}
\end{center}\end{sloppypar}
\vspace{2mm}
\begin{sloppypar}
\noindent
C.K.~Bowdery,
P.G.~Buck,
A.J.~Finch,
F.~Foster,
G.~Hughes,
R.W.L.~Jones,
N.A.~Robertson,
\linebreak
M.I.~Williams
\nopagebreak
\begin{center}
\parbox{15.5cm}{\sl\samepage
Department of Physics, University of Lancaster, Lancaster LA1 4YB,
United Kingdom$^{10}$}
\end{center}\end{sloppypar}
\vspace{2mm}
\begin{sloppypar}
\noindent
I.~Giehl,
K.~Jakobs,
K.~Kleinknecht,
G.~Quast,
B.~Renk,
E.~Rohne,
\mbox{H.-G.~Sander},
H.~Wachsmuth,
C.~Zeitnitz
\nopagebreak
\begin{center}
\parbox{15.5cm}{\sl\samepage
Institut f\"ur Physik, Universit\"at Mainz, D-55099 Mainz, Germany$^{16}$}
\end{center}\end{sloppypar}
\vspace{2mm}
\begin{sloppypar}
\noindent
J.J.~Aubert,
A.~Bonissent,
J.~Carr,
P.~Coyle,
P.~Payre,
D.~Rousseau
\nopagebreak
\begin{center}
\parbox{15.5cm}{\sl\samepage
Centre de Physique des Particules, Facult\'e des Sciences de Luminy,
IN$^{2}$P$^{3}$-CNRS, F-13288 Marseille, France}
\end{center}\end{sloppypar}
\vspace{2mm}
\begin{sloppypar}
\noindent
M.~Aleppo,
M.~Antonelli,
F.~Ragusa
\nopagebreak
\begin{center}
\parbox{15.5cm}{\sl\samepage
Dipartimento di Fisica, Universit\`a di Milano e INFN Sezione di Milano,
I-20133 Milano, Italy}
\end{center}\end{sloppypar}
\vspace{2mm}
\begin{sloppypar}
\noindent
V.~B\"uscher,
H.~Dietl,
G.~Ganis,
K.~H\"uttmann,
G.~L\"utjens,
C.~Mannert,
W.~M\"anner,
\mbox{H.-G.~Moser},
S.~Schael,
R.~Settles,
H.~Seywerd,
H.~Stenzel,
W.~Wiedenmann,
G.~Wolf
\nopagebreak
\begin{center}
\parbox{15.5cm}{\sl\samepage
Max-Planck-Institut f\"ur Physik, Werner-Heisenberg-Institut,
D-80805 M\"unchen, Germany\footnotemark[16]}
\end{center}\end{sloppypar}
\vspace{2mm}
\begin{sloppypar}
\noindent
P.~Azzurri,
J.~Boucrot,
O.~Callot,
S.~Chen,
A.~Cordier,
M.~Davier,
L.~Duflot,
\mbox{J.-F.~Grivaz},
Ph.~Heusse,
A.~Jacholkowska,$^{1}$
F.~Le~Diberder,
J.~Lefran\c{c}ois,
\mbox{A.-M.~Lutz},
\mbox{M.-H.~Schune},
\mbox{J.-J.~Veillet},
I.~Videau,$^{1}$
D.~Zerwas
\nopagebreak
\begin{center}
\parbox{15.5cm}{\sl\samepage
Laboratoire de l'Acc\'el\'erateur Lin\'eaire, Universit\'e de Paris-Sud,
IN$^{2}$P$^{3}$-CNRS, F-91898 Orsay Cedex, France}
\end{center}\end{sloppypar}
\vspace{2mm}
\begin{sloppypar}
\noindent
G.~Bagliesi,
T.~Boccali,
C.~Bozzi,$^{12}$
G.~Calderini,
R.~Dell'Orso,
I.~Ferrante,
L.~Fo\`{a},
A.~Giassi,
A.~Gregorio,
F.~Ligabue,
P.S.~Marrocchesi,
A.~Messineo,
F.~Palla,
G.~Rizzo,
G.~Sanguinetti,
G.~Sguazzoni,
R.~Tenchini,$^{1}$
A.~Venturi,
P.G.~Verdini
\samepage
\begin{center}
\parbox{15.5cm}{\sl\samepage
Dipartimento di Fisica dell'Universit\`a, INFN Sezione di Pisa,
e Scuola Normale Superiore, I-56010 Pisa, Italy}
\end{center}\end{sloppypar}
\vspace{2mm}
\begin{sloppypar}
\noindent
G.A.~Blair,
G.~Cowan,
M.G.~Green,
T.~Medcalf,
J.A.~Strong
\nopagebreak
\begin{center}
\parbox{15.5cm}{\sl\samepage
Department of Physics, Royal Holloway \& Bedford New College,
University of London, Surrey TW20 OEX, United Kingdom$^{10}$}
\end{center}\end{sloppypar}
\vspace{2mm}
\begin{sloppypar}
\noindent
D.R.~Botterill,
R.W.~Clifft,
T.R.~Edgecock,
P.R.~Norton,
J.C.~Thompson,
I.R.~Tomalin
\nopagebreak
\begin{center}
\parbox{15.5cm}{\sl\samepage
Particle Physics Dept., Rutherford Appleton Laboratory,
Chilton, Didcot, Oxon OX11 OQX, United Kingdom$^{10}$}
\end{center}\end{sloppypar}
\vspace{2mm}
\begin{sloppypar}
\noindent
\mbox{B.~Bloch-Devaux},
P.~Colas,
S.~Emery,
W.~Kozanecki,
E.~Lan\c{c}on,
\mbox{M.-C.~Lemaire},
E.~Locci,
P.~Perez,
J.~Rander,
\mbox{J.-F.~Renardy},
A.~Roussarie,
\mbox{J.-P.~Schuller},
J.~Schwindling,
A.~Trabelsi,$^{21}$
B.~Vallage
\nopagebreak
\begin{center}
\parbox{15.5cm}{\sl\samepage
CEA, DAPNIA/Service de Physique des Particules,
CE-Saclay, F-91191 Gif-sur-Yvette Cedex, France$^{17}$}
\end{center}\end{sloppypar}
\vspace{2mm}
\begin{sloppypar}
\noindent
S.N.~Black,
J.H.~Dann,
R.P.~Johnson,
H.Y.~Kim,
N.~Konstantinidis,
A.M.~Litke,
M.A. McNeil,
\linebreak
G.~Taylor
\nopagebreak
\begin{center}
\parbox{15.5cm}{\sl\samepage
Institute for Particle Physics, University of California at
Santa Cruz, Santa Cruz, CA 95064, USA$^{22}$}
\end{center}\end{sloppypar}
\vspace{2mm}
\begin{sloppypar}
\noindent
C.N.~Booth,
S.~Cartwright,
F.~Combley,
M.~Lehto,
L.F.~Thompson
\nopagebreak
\begin{center}
\parbox{15.5cm}{\sl\samepage
Department of Physics, University of Sheffield, Sheffield S3 7RH,
United Kingdom$^{10}$}
\end{center}\end{sloppypar}
\vspace{2mm}
\begin{sloppypar}
\noindent
K.~Affholderbach,
A.~B\"ohrer,
S.~Brandt,
C.~Grupen,
J.~Hess,
A.~Misiejuk,
G.~Prange,
U.~Sieler
\nopagebreak
\begin{center}
\parbox{15.5cm}{\sl\samepage
Fachbereich Physik, Universit\"at Siegen, D-57068 Siegen,
 Germany$^{16}$}
\end{center}\end{sloppypar}
\vspace{2mm}
\begin{sloppypar}
\noindent
G.~Giannini,
B.~Gobbo
\nopagebreak
\begin{center}
\parbox{15.5cm}{\sl\samepage
Dipartimento di Fisica, Universit\`a di Trieste e INFN Sezione di Trieste,
I-34127 Trieste, Italy}
\end{center}\end{sloppypar}
\vspace{2mm}
\begin{sloppypar}
\noindent
J.~Rothberg,
S.~Wasserbaech
\nopagebreak
\begin{center}
\parbox{15.5cm}{\sl\samepage
Experimental Elementary Particle Physics, University of Washington, WA 98195
Seattle, U.S.A.}
\end{center}\end{sloppypar}
\vspace{2mm}
\begin{sloppypar}
\noindent
S.R.~Armstrong,
P.~Elmer,
D.P.S.~Ferguson,
Y.~Gao,
S.~Gonz\'{a}lez,
O.J.~Hayes,
H.~Hu,
S.~Jin,
J.~Kile,
P.A.~McNamara III,
J.~Nielsen,
W.~Orejudos,
Y.B.~Pan,
Y.~Saadi,
I.J.~Scott,
J.~Walsh,
\mbox{J.H.~von~Wimmersperg-Toeller},
Sau~Lan~Wu,
X.~Wu,
G.~Zobernig
\nopagebreak
\begin{center}
\parbox{15.5cm}{\sl\samepage
Department of Physics, University of Wisconsin, Madison, WI 53706,
USA$^{11}$}
\end{center}\end{sloppypar}
}
\footnotetext[1]{Also at CERN, 1211 Geneva 23, Switzerland.}
\footnotetext[2]{Now at Universit\'e de Lausanne, 1015 Lausanne, Switzerland.}
\footnotetext[3]{Also at Centro Siciliano di Fisica Nucleare e Struttura
della Materia, INFN, Sezione di Catania, 95129 Catania, Italy.}
\footnotetext[4]{Also Istituto di Fisica Generale, Universit\`{a} di
Torino, 10125 Torino, Italy.}
\footnotetext[5]{Also Istituto di Cosmo-Geofisica del C.N.R., Torino,
Italy.}
\footnotetext[6]{Supported by the Commission of the European Communities,
contract ERBFMBICT982894.}
\footnotetext[7]{Supported by CICYT, Spain.}
\footnotetext[8]{Supported by the National Science Foundation of China.}
\footnotetext[9]{Supported by the Danish Natural Science Research Council.}
\footnotetext[10]{Supported by the UK Particle Physics and Astronomy Research
Council.}
\footnotetext[11]{Supported by the US Department of Energy, grant
DE-FG0295-ER40896.}
\footnotetext[12]{Now at INFN Sezione de Ferrara, 44100 Ferrara, Italy.}
\footnotetext[13]{Supported by the US Department of Energy, contract
DE-FG05-92ER40742.}
\footnotetext[14]{Supported by the US Department of Energy, contract
DE-FC05-85ER250000.}
\footnotetext[15]{Permanent address: Universitat de Barcelona, 08208 Barcelona,
Spain.}
\footnotetext[16]{Supported by the Bundesministerium f\"ur Bildung,
Wissenschaft, Forschung und Technologie, Germany.}
\footnotetext[17]{Supported by the Direction des Sciences de la
Mati\`ere, C.E.A.}
\footnotetext[18]{Supported by the Austrian Ministry for Science and Transport.}
\footnotetext[19]{Now at SAP AG, 69185 Walldorf, Germany.}
\footnotetext[20]{Now at Harvard University, Cambridge, MA 02138, U.S.A.}
\footnotetext[21]{Now at D\'epartement de Physique, Facult\'e des Sciences de Tunis, 1060 Le Belv\'ed\`ere, Tunisia.}
\footnotetext[22]{Supported by the US Department of Energy,
grant DE-FG03-92ER40689.}
%
%
\setlength{\parskip}{\saveparskip}
\setlength{\textheight}{\savetextheight}
\setlength{\topmargin}{\savetopmargin}
\setlength{\textwidth}{\savetextwidth}
\setlength{\oddsidemargin}{\saveoddsidemargin}
\setlength{\topsep}{\savetopsep}
\normalsize
\newpage
\pagestyle{plain}
\setcounter{page}{1}

%
%
\section{Introduction}

  Quantum chromodynamics (QCD) predicts the existence of 
bound states of gluons, called glueballs.
Lattice calculations have predicted the lightest scalar glueball to be 
a scalar resonance with mass $1600 \pm 150 \mevsq$~\cite{UKQCD,IBM} with
tensor and pseudoscalar glueballs in the $2000-2500 \mevsq$ region~\cite{madname}.

 Experimentally two principal candidates for the scalar glueball have been 
observed, the $f_0(1500)$ and the $f_J(1710)$ (a review is given 
in~\cite{PDG}). 
Both are seen in gluon-rich reactions
such as $p\bar{p}$ annihilation, central production and radiative 
\Jpsi decay. The $f_0(1500)$
does not fit naturally in the $q\bar{q}$ spectrum but could be due to
a glueball mixed with $q\bar{q}$ states in that mass region~\cite{amsler}.

  The spin of the $f_J(1710)$ is not yet confirmed, indications for both spin 2
and spin 0 have been reported~\cite{PDG,spin}. If the $f_J(1710)$ is indeed
spin 0 it becomes a candidate for the $^{3}P_{0}(s\bar{s})$-like state
in the $1600-2000 \mevsq$ region, but also for the lightest scalar glueball~\cite{IBM}.

 In $\gamma \gamma$ interactions, production of a pure gluon state is
suppressed. Measuring the two-photon width $\Gamma_{\gamma \gamma}$ of
the $f_0(1500)$ and the $f_J(1710)$, or setting an upper limit on 
$\Gamma_{\gamma \gamma}$, should indicate whether either is likely to be a pure
glueball or has quark content.
The CLEO collaboration has recently published a stringent limit on the
two-photon width of the $f_{J}(2220)$ resonance (formerly the $\xi(2220)$)
which is a candidate for the tensor glueball~\cite{cleo}.

 At the LEP $e^{+}e^{-}$ collider there is a large cross-section for
inelastic two-photon scattering, in which each of the incoming electrons
acts as a source of virtual photons. In this analysis, the processes
$\gamma \gamma \rightarrow f_0(1500) \rightarrow \pi^+ \pi^-$ and 
$\gamma \gamma \rightarrow f_J(1710) \rightarrow \pi^+ \pi^-$ 
have been studied and upper limits extracted for 
$\Gamma_{\gamma \gamma} \cdot {\cal BR}(\pi^{+}\pi^{-})$ of both resonances.

%
%
\section{The ALEPH detector}

  The ALEPH detector and its performance are described in detail
elsewhere~\cite{aleph,alephperformance}. Here only a brief description 
of the detector components relevant for this analysis is given.

  The trajectories of charged particles are measured with a silicon vertex
detector (VDET), a cylindrical multiwire drift chamber (ITC) and a large time 
projection chamber
(TPC).
The three detectors are immersed in a 1.5 T axial magnetic field from the 
superconducting solenoidal coil and together provide a transverse momentum
resolution 
$\delta p_{\perp}/p_{\perp} = 6 \times 10^{-4}p_{\perp} \oplus$ 0.005
($p_{\perp}$ in \gevc).
The TPC also provides up to 338 measurements of 
ionisation (d{\em E}/d{\em x}) used for particle identification.

Between the TPC and the coil, an electromagnetic calorimeter (ECAL) is used
to identify electrons and photons and to measure their energy with a
relative resolution of 0.18/$\sqrt{E} + 0.009$ ($E$ in GeV). The 
luminosity calorimeters (LCAL and SiCAL) cover the small polar angle
region, 24--190 mrad. 

Muons are identified by their characteristic penetration pattern in 
the hadron calorimeter (HCAL), a 1.2 m thick iron yoke instrumented with 
23 layers of limited streamer tubes, together with two surrounding layers of
muon chambers. In association with the electromagnetic calorimeter, the
hadron calorimeter also provides a measurement of the energy of charged
and neutral hadrons with a relative resolution of 0.85/$\sqrt{E}$ ($E$ in GeV).

 The main ALEPH trigger relevant for this analysis, for data taken in 1994 
and after, is based on the
identification of two track candidates in the ITC,
with at least one track pointing to an energy deposit in excess of 200 MeV
in the ECAL. The two track candidates are confirmed at the second trigger 
level using the TPC. For data taken before 1994,
the analysis relies on a trigger of two ITC (and TPC) track candidates measured
back-to-back within $11^{\circ}$.

%
%
\section{Monte Carlo samples}

  Fully simulated Monte Carlo event samples reconstructed with the
same program as the data have been used for the design of the event
selection, background estimation and extraction of a limit on 
two-photon widths $\Gamma_{\gamma\gamma}^{f_{0}(1500)}$ and 
$\Gamma_{\gamma\gamma}^{f_{J}(1710)}$. Samples of the signal
processes 
$\gamma \gamma \rightarrow f_{0}(1500) \rightarrow \pi^{+} \pi^{-}$ and
$\gamma \gamma \rightarrow f_{J}(1710) \rightarrow \pi^{+} \pi^{-}$
were generated using
PHOT02~\cite{PHOT02}. Each sample was generated as a Breit-Wigner resonance
of appropriate mass, width and spin, with the $f_{J}(1710)$ here assumed
to be spin zero. The experimental resolution of the masses of
both
resonances is of the order of $10\mevsq$, compared to full widths in excess
of $100\mevsq$.

Expected background processes $\gamma \gamma \rightarrow e^{+} e^{-}$,
$\gamma \gamma \rightarrow \mu^{+} \mu^{-}$,
and $\gamma \gamma \rightarrow \tau^{+} \tau^{-}$
were also generated using
PHOT02, while background from 
$e^{+}e^{-} \rightarrow {\rm Z} \rightarrow \tau^{+} \tau^{-}$ was estimated
using KORALZ~\cite{koralz}. Both programs are interfaced to the
JETSET~\cite{jetset} package for hadronisation, while KORALZ also includes
TAUOLA~\cite{tauola} for correct handling of the $\tau$ polarisation.

%
%
\section{Event selection}

This analysis uses 160.9 \pb of data taken around $\sqrt{s}=91$ GeV
from 1990 to 1995.
 Candidate events for $\gamma \gamma \rightarrow \pi^{+} \pi^{-}$
are selected according to the following criteria:
\begin{itemize}
 \item  the event contains only two good tracks, of equal and opposite 
  charge. Good tracks are defined to have at least four TPC hits, 
  ${|}\cos\theta{|} < 0.93$, where $\theta$ is the polar angle with respect to 
beam axis, and a minimum distance to the interaction
  point of less than 8 cm along the beam axis and 2 cm in the radial direction;
 \item the total energy summed over all objects reconstructed by the
energy flow program is equal to the total energy of the
two charged tracks;
 \item  the primary vertex is reconstructed within $\pm 2$ cm of the nominal
interaction point along the beam axis;
 \item  the total energy observed in the event is less than 30 GeV and 
visible mass less than 10 GeV, to exclude Z events;
 \item  the transverse momentum of the final state with respect to the
beam axis does not exceed $0.1 \gevc$;
\item  the absolute value of the cosine of the angle $\theta^*$ 
between the two 
tracks in their centre-of-mass system and the boost direction is required to
be less than 0.9.
\end{itemize}

 These criteria ensure (quasi-)real photon ($Q^{2}=0$) collisions, and also
suppress three- or more body decay.


The dominant background to the two-pion signal is the 
$\gamma \gamma \rightarrow \mu^+ \mu^-$ process. 
At low energies, 
muons and pions cannot easily be distinguished from each other and the 
high cross-section for 
the  $\gamma \gamma \rightarrow \mu^+ \mu^-$ process leads to a
substantial contamination of the $\pi^+ \pi^-$ sample. 

To reduce background, events containing identified 
muons~\cite{alephperformance} are rejected. With a 
sample of $\gamma \gamma \rightarrow \mu^+ \mu^-$ events generated 
using PHOT02, it was found 
that the efficiency of rejection is about 100\%
for events with invariant mass $W>3\gevsq$. 
Below $W=3\gevsq$, the rejection efficiency is improved by means of 
simple cuts on the fraction of energy deposited in the ECAL and HCAL, 
but still the rejection efficiency falls off 
rapidly, being only ${\sim} 20$\% for $W=1\gevsq$. For $W\geq 1\gevsq$ 
the average rejection efficiency for
$\gamma \gamma \rightarrow \mu^+ \mu^-$ events is 45\%.

After muon rejection, the specific ionisation (d{\em E}/d{\em x}) of 
each track, 
measured in the TPC, is required to be within three standard deviations
of the expected ionisation for a pion and more than three standard deviations 
from that of an electron.
Separation of pions and kaons by \dedx is less efficient, but the rate of 
kaon pair production is low: from
a Monte Carlo sample of $f^{\prime}_{2}(1525) \rightarrow K^{+}K^{-}$, a 
residual contamination of only ${\sim}0.05\%$ was estimated.
  
Possible background from beam-gas interactions was investigated. Constraining
the reconstructed primary vertex position in $z$ removes most of this 
background. As beam-gas events are uniformly distributed in $z$,
the number of events falling outside the $z$-vertex cut can be used 
to estimate 
the remaining background inside the cut: beam-gas contamination is found 
to be negligible. 
Residual backgrounds from the processes
$\gamma \gamma \rightarrow \tau^+ \tau^-$ and
$Z \rightarrow \tau^+ \tau^-$ are also negligible. 

 A sample of 294141 $\pi^{+}\pi^{-}$ candidate events is selected.
%
%
\section{Fitting the mass spectrum}

\begin{figure}[htbp]
\vspace{-5mm}
\epsfxsize=10cm
\epsfysize=7.5cm
\begin{center}\mbox{\epsfbox{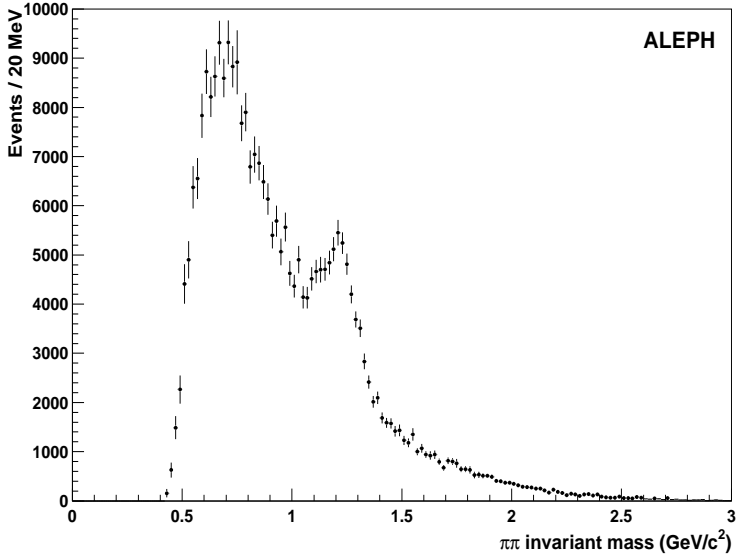}}\end{center}
\caption{\label{pipimass} \footnotesize
The invariant mass distribution for two-pion final states. The error bars
indicate statistical errors only.}
\end{figure}

  The $\pi^+ \pi^-$ invariant mass spectrum obtained is shown in 
Fig.~\ref{pipimass}. The steep rise of the spectrum around $0.5\gevsq$
is an artefact of the trigger efficiency. The clear peak in the spectrum 
above $1\gevsq$ can be identified as the known tensor resonance $f_{2}(1270)$.
No other structure is observed above the $f_{2}(1270)$ peak.

 A fit to the invariant mass spectrum is performed. For the resonances 
$f_{2}(1270)$, $f_{0}(1500)$ and $f_{J}(1710)$,
a Breit-Wigner function of the form
\begin{equation} \label{bw} \frac{m m_{0} \Gamma(m)}{(m_{0}^{2} - m^{2})^{2}
 + m^{2} \Gamma^{2}(m)} \end{equation} 
is used, where 
the mass-dependent width $\Gamma(m)$, away from the two-pion threshold, 
has the form
\begin{equation} \label{width} \Gamma(m) = \Gamma_{0}      
\left(\frac{m}{m_{0}}\right)^{(l+1/2)}\exp\left[
\frac{-(m^{2}-m_{0}^{2})}{48 \beta^2}\right]
\end{equation}
and $\beta=0.4\gevsq$~\cite{sandy}.
For the $f_{2}(1270)$ the mass $m_{0}$, total width $\Gamma_{0}$ 
and overall normalisation are free parameters. For the $f_{0}(1500)$ and 
the $f_{J}(1710)$ the normalisation is a free parameter, while the mass
and width are fixed to $1500\mevsq$ and $112\mevsq$ for the
$f_{0}(1500)$ and $1712\mevsq$ and $133\mevsq$ for the $f_{J}(1710)$~\cite{PDG}. 
The $f_{J}(1710)$ is considered here as a $J=0$ state. Although the presence of
two objects in the $1700\mevsq$ region has been suggested~\cite{godknows}, no 
attempt to resolve two objects is made here.

The background spectrum, due to 
$\gamma \gamma \rightarrow \mu \mu$ events and
the $\pi\pi$ continuum process, is fitted on the data with a $5^{th}$ order 
Chebyshev polynomial.

  The mass region 0.8 to $2.5\gevsq$ is used in the fit (below $0.8\gevsq$ the
trigger efficiency falls to less than 20\%). Four fits are performed,
each including a Breit-Wigner for the $f_{2}(1270)$ and
the polynomial background and then including \vspace{-2mm}
\begin{itemize}
\item[(i)] no additional resonances, \vspace{-3mm}
\item[(ii)] the $f_{0}(1500)$,\vspace{-3mm}
\item[(iii)] the $f_{J}(1710)$,\vspace{-3mm}
\item[(iv)] both the $f_{0}(1500)$ and the $f_{J}(1710)$.\vspace{-2mm}
\end{itemize}
\begin{figure}[htbp]
\epsfxsize=10cm
\epsfysize=7.5cm
\begin{center}\mbox{\epsfbox{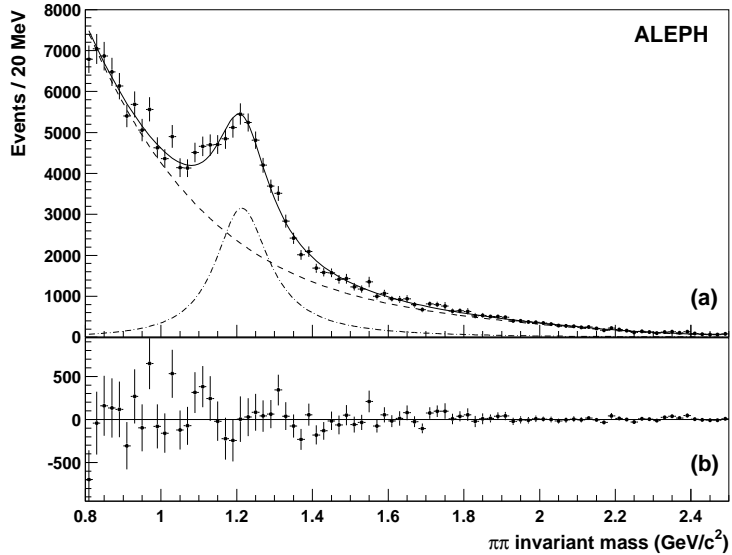}}\end{center}
\caption{\label{fitmass} \footnotesize
(a) The fit to data with a Breit-Wigner function for the $f_{2}(1270)$
(dot-dash line), a polynomial for the background (dashed line) and
the combination of these functions (solid line); and 
(b) the data after subtraction of the fitted curve. Error bars indicate
statistical errors only.
}
\end{figure}
 The fit to the data is shown in Fig.~\ref{fitmass} for case (i): 
the Breit-Wigner shape for the $f_{2}(1270)$, the polynomial for background
processes and the sum of these are indicated.

The parameters 
of all fits are summarised in Table~\ref{fitparam}, including the number
of events fitted for the glueball candidate signals. For the fit including the
$f_{2}(1270)$ alone, the $\chi^2$ of 75 for 76 degrees of freedom 
is already very good.
The $\chi^2$ per degree of freedom hardly changes with the addition of
glueball signals. The fits including the $f_{J}(1710)$ were also performed
for $J=2$. The results of these fits were essentially the same as for the 
$J=0$ case.

\begin{table}[htbp]
\caption{\label{fitparam} \footnotesize Table of parameters for fits
 using the Breit-Wigner form of Eqn.\ref{bw} for resonances.}
\begin{center}
\begin{tabular}{|l|c|}\hline\hline
 & (i) Fit for $f_{2}(1270)$ only \\
\hline
$\chi^2$           & 74.96 \\
Degrees of freedom & 76 \\
Fitted mass of $f_{2}(1270)$ ($\mevsq$)  & $1213.5 \pm 3.7$\\  
Fitted width of $f_{2}(1270)$ ($\mevsq$) & $178.3 \pm 12.8$\\
\hline
 & (ii) Fit for $f_{2}(1270) + f_{0}(1500)$ \\
\hline
$\chi^2$           & 73.33 \\
Degrees of freedom & 75 \\
Fitted mass of $f_{2}(1270)$ ($\mevsq$)  & $1214.1 \pm 3.8$\\  
Fitted width of $f_{2}(1270)$ ($\mevsq$) & $173.9 \pm 13.8$\\
No. of $f_{0}(1500)$ signal events & $-808.3 \pm 602.6$ \\
\hline
 & (iii) Fit for $f_{2}(1270) + f_{J}(1710)$ \\
\hline
$\chi^2$           & 74.02 \\
Degrees of freedom & 75 \\
Fitted mass of $f_{2}(1270)$ ($\mevsq$)  & $1213.9 \pm 3.9$\\  
Fitted width of $f_{2}(1270)$ ($\mevsq$)  & $180.2 \pm 15.9$\\
No. of $f_{J}(1710)$ signal events & $468.3 \pm 476.6$ \\
\hline
 & (iv) Fit for $f_{2}(1270) + f_{0}(1500) + f_{J}(1710)$ \\
\hline
$\chi^2$           & 73.21 \\
Degrees of freedom & 74 \\
Fitted mass of $f_{2}(1270)$ ($\mevsq$)  & $1214.2 \pm 3.8$\\  
Fitted width of $f_{2}(1270)$ ($\mevsq$)  & $175.5 \pm 14.2$\\
No. of $f_{0}(1500)$ signal events & $-671.0 \pm 690.0$ \\
No. of $f_{J}(1710)$ signal events & $198.6 \pm 541.9$ \\
\hline\hline
\end{tabular}
\end{center}
\end{table} 

  The width of the Breit-Wigner function fitted for the $f_{2}(1270)$ is
in all cases in agreement with the world average value of 
$185.5^{+3.8}_{-2.7}\mevsq$~\cite{PDG}. The fitted mass of ${\sim}1214\mevsq$
in each case is not consistent with the established 
value of $1275.0 \pm 1.2\mevsq$~\cite{PDG}, 
however. This has been previously observed by the MARKII and
CELLO collaborations~\cite{f2shift} and is believed to be caused by 
interference of the spin 2 resonant amplitude with other components in the 
background.

Limited knowledge of the trigger efficiency for this topology prevents
an investigation of interference effects using the measured angular 
distribution.
The number of events fitted for the $f_{0}(1500)$ signal is negative, but consistent with zero. Data
from the WA76 and WA102 collaborations~\cite{WA102} and recent 
calculations~\cite{anisovich} suggest 
interference 
effects in this region can cause the $f_{0}(1500)$ to appear as a dip in the 
spectrum. However, here limits are set assuming no interference.

\begin{figure}[htbp]
\epsfxsize=10cm
\epsfysize=10cm
\begin{center}\mbox{\epsfbox{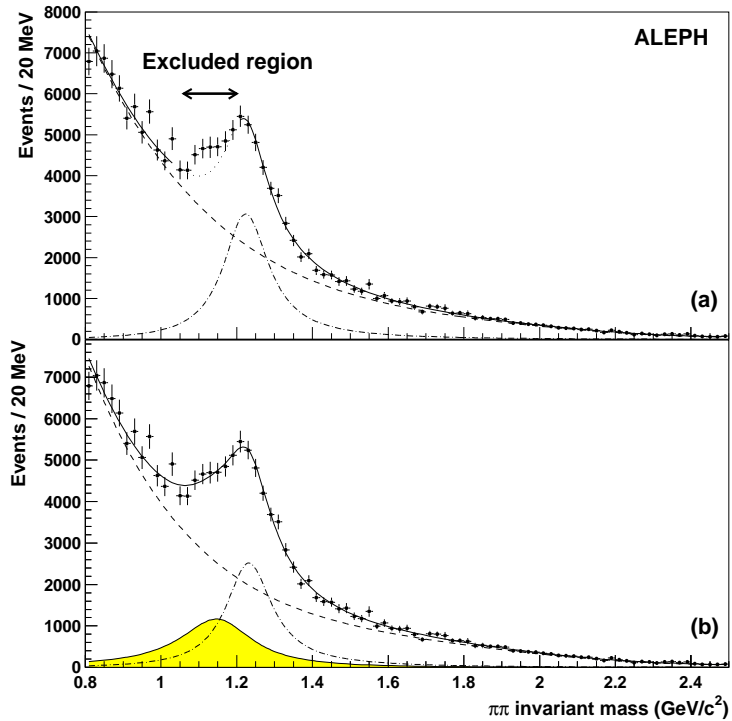}}\end{center}
\caption{\label{fits0} \footnotesize
 The fit to data is shown 
with a Breit-Wigner function for the $f_{2}(1270)$
(dot-dash line), a polynomial for the background (dashed line) and
the sum of these functions (solid line), where 
(a) the region around $1.1\gevsq$ is excluded from the fit --
 the extrapolation of the fit through the excluded region is indicated by 
 the dotted line; and
(b) a second Breit-Wigner function has been introduced around 
$1.1\gevsq$ (shaded 
 region) and is included in the total fit (solid line).
All error bars indicate statistical errors only.
}
\end{figure}

The CELLO collaboration has reported possible structure in the $\pi^{+}\pi^{-}$
invariant mass spectrum around $1.1\gevsq$~\cite{CELLO}. 
If the fit of case (i) is repeated but excluding a region around
$1.1\gevsq$, a clear excess of data over the fitted curve extrapolated 
through that
region can be seen (Fig.~\ref{fits0}a). The excess can be described
by the introduction of another spin 0 resonance of apparent mass 
${\sim}1.1\gevsq$
and total width ${\sim}250\mevsq$ (Fig.~\ref{fits0}b).  

%
%


\section{{\boldmath $\Gamma_{\gamma \gamma} \cdot {\cal BR}(\pi^{+}\pi^{-})$}
 for 
{\boldmath $f_{0}(1500)$} and {\boldmath $f_{J}(1710)$}}

   The fitted numbers of signal events from the processes
$\gamma \gamma \rightarrow f_{0}(1500)\rightarrow \pi^{+}\pi^{-}$ and 
$\gamma \gamma \rightarrow f_{J}(1710)\rightarrow \pi^{+}\pi^{-}$ 
are used to calculate upper limits on $\Gamma_{\gamma \gamma} \cdot {\cal BR}(\pi^{+}\pi^{-})$ for both resonances.
 
The trigger efficiency is calculated as a function of invariant mass
by comparing rates of independent 
triggers for $\pi^{+}\pi^{-}$ selected data. In the $1500\mevsq$ 
mass region the average trigger efficiency is 
($66 \pm 7$)\%, while for the $1700\mevsq$ mass region it is ($77 \pm 7$)\%.


The acceptance and selection efficiency for $\gamma \gamma \rightarrow 
f_0(1500) \rightarrow \pi^+ \pi^-$ and for $\gamma \gamma \rightarrow 
f_J(1710) \rightarrow \pi^+ \pi^-$ are determined from Monte Carlo
to be ($17.5 \pm 0.4$)\% and ($16.3 \pm 0.4$)\%, where the quoted 
errors include
the systematic error due to the simulation of the detector (checked by varying
the resolution on measured quantities used in the event selection) and
the error due to Monte Carlo statistics.

 The product branching ratio, 
$\Gamma_{\gamma \gamma} \cdot {\cal BR}(\pi^{+}\pi^{-})$,
 of each resonance is calculated directly from the
fitted number of signal events in each of the cases (ii)--(iv). The fitted 
value of $\Gamma_{\gamma \gamma} \cdot {\cal BR}(\pi^{+}\pi^{-})$ (which may 
be negative) and its error are used to define 
the mean and width of a Gaussian 
distribution: the area under the Gaussian is integrated above zero to obtain
the 95\% \CL limit on $\Gamma_{\gamma \gamma} \cdot {\cal BR}(\pi^{+}\pi^{-})$.

The upper limits on the product branching ratios at 95\% \CL are,
for the individual fits, 
$\Gamma_{\gamma \gamma} \cdot {\cal BR}(\pi^{+}\pi^{-}) < 0.25$ keV and
$\Gamma_{\gamma \gamma} \cdot {\cal BR}(\pi^{+}\pi^{-}) < 0.59$ keV for the
$f_{0}(1500)$ and $f_{J}(1710)$ respectively, and for the combined fit are
$\Gamma_{\gamma \gamma} \cdot {\cal BR}(\pi^{+}\pi^{-}) < 0.31$ keV and
$\Gamma_{\gamma \gamma} \cdot {\cal BR}(\pi^{+}\pi^{-}) < 0.55$ keV for the
$f_{0}(1500)$ and $f_{J}(1710)$ respectively.
If an additional spin 0 resonance with a fixed mass of $1.1\gevsq$ (as 
discussed in section 5) is included in the combined fit, the limit on the 
product branching ratio for the $f_{0}(1500)$ worsens by ${\sim} 30\%$ and 
for the $f_{J}(1710)$ tightens by ${\sim} 10\%$.



 The analysis was repeated using 
the same  Breit-Wigner form for the resonances but with an alternative 
expression for the mass-dependent width that has been used in some
previous experimental resonance studies~\cite{omega}:
\begin{equation}\Gamma(m) = \Gamma_{0}   \cdot
\frac{2 \cdot (
\frac{m^{2} - 4 m_{\pi}^{2}}{m_{0}^{2} - 4 m_{\pi}^{2}}
)^{(l+\frac{1}{2})}}
{1+\frac{m^{2} - 4 m_{\pi}^{2}}{m_{0}^{2} - 4 m_{\pi}^{2}}}.
\end{equation}

The fitted parameters and numbers of signal events are given in 
Table~\ref{fitomega}. The $\chi^2$ per degree of freedom for each of these fits
is considerably worse than for the original fits.

\begin{table}[htbp]
\caption{\label{fitomega} \footnotesize Table of parameters for fits using an alternative expression for the mass-dependent width of all resonances.}
\begin{center}
\begin{tabular}{|l|c|}\hline\hline
 & (i) Fit for $f_{2}(1270)$ only \\
\hline
$\chi^2$           & 108.36 \\
Degrees of freedom & 76 \\
Fitted mass of $f_{2}(1270)$ ($\mevsq$)  & $1216.2 \pm 18.5$\\  
Fitted width of $f_{2}(1270)$ ($\mevsq$) & $221.3 \pm 18.2$\\
\hline
 & (ii) Fit for $f_{2}(1270) + f_{0}(1500)$ \\
\hline
$\chi^2$           & 88.48 \\
Degrees of freedom & 75 \\
No. of $f_{0}(1500)$ signal events & $-1549.4 \pm 901.9$ \\
\hline
 & (iii) Fit for $f_{2}(1270) + f_{J}(1710)$ \\
\hline
$\chi^2$           & 89.81 \\
Degrees of freedom & 75 \\
No. of $f_{J}(1710)$ signal events & $1019.8 \pm 475.6$ \\
\hline
 & (iv) Fit for $f_{2}(1270) + f_{0}(1500) + f_{J}(1710)$ \\
\hline
$\chi^2$           & 87.51 \\
Degrees of freedom & 74 \\
No. of $f_{0}(1500)$ signal events & $-1242.8 \pm 741.1$ \\
No. of $f_{J}(1710)$ signal events & $485.2 \pm 576.0$ \\
\hline\hline
\end{tabular}
\end{center}
\end{table}

The upper limit on $\Gamma_{\gamma\gamma} \cdot {\cal BR}(\pi^{+}\pi^{-})$ 
for the case when only the
$f_0(1500)$ is included is 0.33 keV, and for the $f_J(1710)$ only is
0.84 keV. For the case including both resonances, the upper limits
are 0.28 keV and 0.69 keV for the $f_0(1500)$ and the $f_J(1710)$
respectively. These limits are consistent with the original results.



\section{Stickiness}

   From limits on 
$\Gamma_{\gamma \gamma}$  the 
{\em stickiness}~\cite{sticky}
of each resonance can be calculated, by comparison of the two photon 
width with the
particle's width for production in the glue-rich environment of radiative 
\Jpsi decay. The ratio is normalised such that the $q\bar{q}$ resonance
$f_{2}(1270)$ has stickiness equal to 1. Then for a glueball a higher
value of stickiness is expected.

Stickiness for a $0^{++}$ resonance $X$ is defined as
$$ S_{X}(0^{++}) = {\cal N} \: \frac{m_X}{k_{{\rm J}/\psi \rightarrow \gamma X}}
\: \frac{\Gamma (\Jpsi \rightarrow \gamma X)}{\Gamma (X \rightarrow \gamma
\gamma)} $$
where $m_X$ is the mass of the resonance;
$k_{{\rm J}/\psi \rightarrow \gamma X}=
(m_{{\rm J}/\psi}^{2} - m_{X}^{2})/2m_{{\rm J}/\psi}$ is the 
energy of the photon from the radiative \Jpsi decay, measured in the
\Jpsi rest frame;
$\Gamma (\Jpsi \rightarrow \gamma X)$ is the width for production of the 
resonance in radiative \Jpsi decay;
and ${\cal N}$ is the normalisation factor.
The branching ratio for $f_{0}(1500) \rightarrow \pi^+ \pi^-$ is 
taken as $0.30 \pm 0.07$, while the branching ratio for 
$f_{J}(1710) \rightarrow \pi^+ \pi^-$ is $0.026^{+0.001}_{-0.016}$~\cite{PDG},
giving, from the fit for both resonances,
$\Gamma (f_{0}(1500) \rightarrow \gamma \gamma)$ = 1.08 keV and
$\Gamma (f_{J}(1710) \rightarrow \gamma \gamma)$ = 21.25 keV.
The lower limits on stickiness are then 1.4 and 0.3 for 
the $f_{0}(1500)$ and 
the $f_{J}(1710)$ respectively.
 

\section{Conclusion}

  Production of the glueball candidates $f_{0}(1500)$ and 
$f_{J}(1710)$ in 
$\gamma \gamma$ collisions at LEP1 has been studied via 
decay to $\pi^+ \pi^-$. No signal from either resonance is seen,
and the upper limits on the product of two-photon width and $\pi^{+}\pi^{-}$
branching ratio have been calculated
as 
$\Gamma_{\gamma \gamma} \cdot {\cal BR}(\pi^{+}\pi^{-}) < 0.31$ keV for the
${f_{0}(1500)}$ and
$\Gamma_{\gamma \gamma} \cdot {\cal BR}(\pi^{+}\pi^{-}) < 0.55$ keV for the
${f_{J}(1710)}$, both at 95\% \CL\hspace{-1.5mm}, from 
a simultaneous fit for both resonances. 

\section*{Acknowledgements}
 We wish to thank our colleagues from the accelerator divisions for
the successful operation of LEP. We are indebted to the engineers and
technicians at CERN and our home institutes for their contribution
to the good performance of ALEPH. Those of us from non-member countries
thank CERN for its hospitality.
%
%

\end{document}